\documentclass[12pt]{article} 
\usepackage{amssymb}
\usepackage{graphicx}
\begin{document} 
\begin{center}
{\large \bf Geometry of ultraperipheral nuclear collisions}

\vspace{0.5cm}                   

{\bf I.M. Dremin}

\vspace{0.5cm}                       

         Lebedev Physical Institute, Moscow 119991, Russia\\

\end{center}


\begin{abstract}
It is advocated that geometry of the interaction region of two heavy
nuclei colliding at large impact parameters is important for the relative
role of light-by-light scattering and QCD-initiated processes.
Exclusive production of resonances is possible by dense electromagnetic
fields in the interior space between the nuclei. The cross section 
of the two-photon processes is evaluated and some examples are considered. 
It is speculated that the exclusive production of $\rho ^0$-mesons by 
"two-photon" processes forbidden by the Landau-Yang rule may be allowed 
within strong magnetic fields due to odd number of photons becoming involved. 
\end{abstract}

\section{Introduction}

Geometry of the internal structure of hadrons and their interaction regions
is of utmost importance for understanding the outcome of these interactions.
Since Yukawa times, the effective size of hadrons is considered being of the
order of 1 fm=$10^{-13}$ cm i.e. close to the inverse pion mass. This scale is 
characteristic for the QCD potential and for simplified bag-models. More tiny 
details about the hadron structure were recently found \cite{beg, ps, fss, dr1}.

High energy interactions of heavy ions open a new avenue for studies of
hadron properties. Central collisions are claimed to produce a new deconfined 
state of QCD-matter named the quark-gluon plasma. In their turn, 
ultraperipheral collisions provide access to strong electromagnetic fields
created within the space between the colliding ions. Exclusive production
of hadronic resonances via light-by-light scattering can become compatible with
QCD sources \cite{hkr}. The geometrical factor plays an important role there. 
Exclusive production of $\rho ^0$-mesons asks for special attention 
because it can happen not only due to the direct transition
of a single photon to $\rho ^0$ in the QCD-field of a nucleus but also
via two-photon collisions in a strong magnetic field. The analogy to creation
of two photons by a single one considered in astrophysics \cite{adl, bms, as}
is useful there.

\section{Exclusive production of resonances in ultraperipheral collisions}

By definition, collisions of two heavy ions are called ultraperipheral if
the shortest distance between the trajectories of their centers (the impact
parameter $b$) exceeds noticeably the sum of their radii 2$R_A$ ($\approx 14$ fm
for Pb-Pb collisions). Unfortunately, the impact parameter can not be directly
measured in experiment. One has to select ultraperipheral events according to
special signatures left by products of the collision. For example, low
multiplicities of final states favor this choice. In particular, exclusive
production of resonances separated by large rapidity gaps from the colliding
ions is a clear indication on ultraperipheral collisions. This process is 
considered as a good candidate for studying various geometrical and dynamical 
aspects of heavy ion interactions.

At large impact parameters the direct short-range interaction of hadronic
constituents (quarks, gluons) of ions occupies small space volume. The
long-range electromagnetic forces have a chance  to dominate because the
electromagnetic fields between two high energy ions are extremely dense. Each 
ion is surrounded by a widely spread cloud of virtual photons. The nuclei emit
the photons coherently. The effective perturbation parameter associated with
photon exchange is not $\alpha $=1/137  but becomes equal $Z\alpha (\approx 0.6$
for Pb ions). The incoherent "inelastic" emission is a factor $Z^{-1}$ smaller
at large charge of the heavy ion $Z$.

The coherent photon flux emitted by a heavy ion has been evaluated a long time 
ago \cite{bgms}. The equivalent photon approximation has been used. 
Coherent emission constitutes the dominant contribution to the photon PDF at
the starting scale of photon virtuality. Impact parameters larger than 19 fm 
are necessary for formation of dense enough electromagnetic fields in Pb-Pb 
collisions at LHC energies (see Fig. 4 in \cite{hkr}) not to have an
additional soft QCD interaction which fills the rapidity gap and destroys
the 'exclusivity' condition. Thus the ions must be
at least 5 fm apart to form the dense electromagnetic field.  These
collisions are really ultraperipheral.

The flux is dominated by photons carrying small fractions of the nucleon 
energy $x$. It may be written \cite{bgms, kn, kmr} as
\begin{equation}
\frac {dn}{dx}=\frac {2Z^2\alpha }{\pi x}\ln \frac {1}{4R_Axm},
\label{e1}
\end{equation}
where the Pb-radius is $R_A$=7 fm and $m$ is a nucleon mass.

The photons in the flux fluctuate to quark-antiquark pairs. They may form
$\rho ^0$-meson when the transverse separation of the pair becomes of the 
hadronic size. Then the $\rho ^0$-meson may elastically scatter on a nucleus.
The strong short-range QCD-forces get involved. Namely they initiate
the diffractive structure of $p_t$-distributions of mesons with 
its characteristic maxima and minima if the black disk model of 
ions is used as done in Ref. \cite{kmr}. This process is semiperipheral
because the target nucleus has to participate in the interaction with $\rho$.  
The transverse momentum spread of initial photons (integrated over in 
Eq. (\ref{e1})) is less than 70 MeV and, by itself, would not influence the 
diffractive minima placed at larger transverse momenta. However, in combination 
with the somewhat larger $p_t$-spread of pomerons this structure
can become strongly smoothed (see Fig. 1 in \cite{kn}\footnote{I am
grateful to S. Klein for this reference.}). Also,
the incoherent and nucleon-break up processes may spoil the diffraction
pattern as shown in Fig. 7 of \cite{kmr}. Some other effects which may smooth
out the diffraction picture are also discussed in \cite{kmr}. 

Indeed, it would be fascinating to observe the well pronounced diffraction
in nuclear collisions at LHC energies. Up to now, a very
smooth behavior of the $p_t$-distribution of exclusively produced 
$\rho $-mesons with slight twists in place of strong diffraction minima 
predicted by Monte Carlo calculations was observed by STAR Collaboration 
(see Fig. 2 in \cite{deb}, also \cite{star, kgs}). Moreover, no distinct 
diffractive structure is
observed nowadays in the data for elastic scattering of hadrons
\cite{totem1, dr2}. Instead, the two exponentially (in $p_t^2$) decreasing 
regions with different exponents and some dip between them are seen. 
The dip is prescribed to the vanishing imaginary part of the
elastic scattering amplitude.

The ultraperipheral processes are those with purely electromagnetic 
interactions of fields between the colliding nuclei. Further background to 
the one-photon mechanism of $\rho ^0$-production considered in Ref. \cite{kmr} 
may come from them. No diffraction pattern of $p_t$ distributions is expected 
to arise in such processes. Recently, elastic scattering
of two photons from proton clouds was observed by ATLAS and CMS Collaborations 
at LHC \cite{atl, cms} even for proton-proton collisions at 13 TeV.
The electromagnetic fields in ultraperipheral collisions of heavy ions are 
almost eight orders of magnitude stronger than those for $pp$-collisions
being enhanced by the $Z^4$-factor. The photon flux (\ref{e1}) becomes very
dense ($dn/dx \approx 300$) for extremely small effective values of 
$x\sim 3\cdot 10^{-4}$ typical for resonance production at LHC energy 
(per nucleon) $\sqrt {s_{nn}}=2.76$ TeV \cite{kmr}. The density of fields
is larger at higher energies because $x=m_R/\sqrt s_{nn}$ is inverse 
proportional to $\sqrt s_{nn}$ for a given resonance $R$.
The hadronic resonances may be copiously created in purely photon
interactions as advocated in many papers (see, e.g., \cite{bgms, hkr} and 
references therein). Geometry of ultraperipheral collisions with the large 
space (more than 5 fm) between the colliding ions and $Z$-factors 
ascribed to both ions favor these reactions. To compare, strong interactions of 
hadronic constituents are limited by distances of the order of 1 fm. 

One can consider the two- and multi-photon processes. Those with the even(odd) 
number of participating photons may produce para(ortho)-states of resonances. 
The transverse momenta are extremely small and $p_t$ distribution of 
produced resonances should be smooth because no diffraction gets
involved in these processes.

As an example, let us consider a particular channel of production
of a resonance $R$ by the two-photon interactions in the ultraperipheral
collisions of Pb-ions at high energies. The exclusive cross section can
be written as
\begin{equation}
\sigma _{AA}(R)=\int dx_1dx_2\frac {dn}{dx_1}\frac {dn}{dx_2}
\sigma _{\gamma \gamma }(R),
\label{e2}
\end{equation}
where $dn/dx_i$ are given by Eq. (\ref{e1}) and (see Ref. \cite{bgms})
\begin{equation}
\sigma _{\gamma \gamma }(R)=\frac {8\pi ^2\Gamma _{tot}(R)}{m _R}
Br(R\rightarrow \gamma \gamma )Br_d(R)\delta (x_1x_2s_{nn}-m_R^2).
\label{e3}
\end{equation}
Here $m_R$ is the mass of $R$, $\Gamma _{tot}(R)$ its total width and  
$Br_d(R)$ denotes the branching ratio to a considered channel of its decay.
The $\delta $-function approximation is used for resonances with small
widths compared to their masses.

The integrals in Eq. (\ref{e2}) can be easily calculated so that one gets
the analytical formula
\begin{equation}
\sigma _{AA}(R)=\frac {128}{3}Z^4\alpha ^2Br(R\rightarrow \gamma \gamma )Br_d(R)
\frac {\Gamma _{tot}(R)}{m_R^3}\ln ^3\frac {\sqrt {s_{nn}}}{4R_Amm_R}.
\label{e4}
\end{equation}
The most fascinating feature of this result is the rather fast increase
of the cross section with increasing energy as $\ln ^3s$. Such energy 
dependence for ultraperipheral collisions was found a long time ago
\cite{ll} but exploited rather recently (see e.g. \cite{vyzh}). The growth
of the flux density (\ref{e1}) is its main source. It overshoots the 
famous Froissart bound $\ln ^2s$ for increase of hadronic cross sections.
The long-range electromagnetic forces admit such a possibility.

Surely, Eq. (\ref{e4}) can be only valid at rather high energies 
\begin{equation}
\sqrt {s_{nn}}\gg 4R_Amm_R\approx 138m_R. 
\label{e5}
\end{equation}
That sets the energy threshold for resonance production in ultraperipheral 
nuclear collisions. It is well satisfied at LHC energies, while RHIC energies
are close to the threshold and NICA's are below it. Let us note the higher
thresholds for heavier resonances in Eq. (\ref{e5}) and smaller cross sections
in Eq. (\ref{e4}).

The simplest example would be to estimate the cross section of exclusive
production of $\pi ^0$-mesons\footnote{Actually, due to problems with measuring 
the total cross section, all estimates may be considered as estimates by an 
order of magnitude for the rapidity distribution at $Y=0$ in the colliding mode. 
Specific features of a particular experimental installation should be
Monte-Carlo accounted.}. In fact, this is the estimate of the lower bound
for light-by-light scattering in ultraperipheral nuclear collisions.
Other non-resonant quark loops would contribute as well beside
the $\gamma \gamma $-state favored by $\pi ^0$.
Even though the $\pi ^0$-lifetime is extremely short 
($\Gamma _{tot}= 7.5\cdot 10^{-6}$ MeV), its rather low mass
($m_{\pi ^0}=135$ MeV) and high branchings equal 1 save the situation.
According to Eq. (\ref{e4}) one gets the values $\sigma _{AA}(R)=15.2$ mb at 
$\sqrt {s_{nn}}=2.76$ TeV and 22 mb at 5.02 TeV, rather optimistic for 
experimentation even in view of some fiducial cuts. 

Another interesting process is the exclusive production of $\eta '$-mesons.
It is especially attractive because
it leads to the additional supply of $\rho ^0$-mesons due to the decay 
$\eta '\rightarrow \rho ^0+\gamma $. The transverse momentum of $\rho ^0$ from
such decay may be as large as 170 MeV. The oscillations at $p_t\sim $120 MeV
predicted in Ref. \cite{kmr} would be filled in if the corresponding cross 
section is large enough. It is estimated from Eq. (\ref{e4}) using the mass, 
width and branching ratios of $\eta '$-meson ($m_{\eta '}=957.8$ MeV, 
$\Gamma _{tot}=0.196$ MeV, Br$_{\gamma \gamma}$=0.0222, 
Br$_{\rho ^0\gamma}$=0.29) from the Particle Data Group tables \cite{pdg}.
As expected, the contribution of this particular channel is rather low,  
about 1.6 mb at 2.76 TeV and 2.8 mb at 5.02 TeV compared to the values 
of hundreds millibarns measured by STAR Collaboration \cite{deb} and predicted 
in Ref.\cite{kmr} for single-photon processes with the partial participation 
of strong interactions. This is mainly due to the low total width of the 
resonance and small 2$\gamma$-branching ratio of $\eta '$ i.e. low effectiveness 
of its production in photon-photon collisions. Contribution of the non-resonant
electromagnetic background and other resonant parastates ($f^0$ etc) is 
expected to be similar and must be Monte-Carlo-computed with account of 
experimental requirements. 

Actually, there could be other sources of $\rho ^0$ production. Very intriguing 
process would be the direct production of $\rho ^0$-mesons (or other orthostates)
by the two-photon collisions within the strong magnetic fields between the
colliding ions. According to some estimates \cite{khar}, the magnetic field
can be as large as $eB\approx (10 - 15)m_{\pi }^2 (\approx 10^{18}$G).
Therefore the photon-photon interactions in such fields can probably become
more effective than the direct interaction with the target nucleus of the 
$\rho ^0$-meson produced by a single exchanged photon, as considered in Ref. 
\cite{kmr}.  Such a contribution may be noticeable at very small transverse 
momenta ($p_t<1/R_A$). The impact of the strong magnetic field
should be considered as the collective effect of the odd number of photons
on a colliding pair of them. As such, it does not violate the Landau-Yang rule.
Unfortunately, we have no methods to estimate the outcome besides considering
some simplest loops of electrons or quarks immersed in this field.
This process reminds the creation of two photons by a single
one passing through the magnetic field considered in astrophysics
\cite{adl, bms, as}. Again, no diffraction patterns in $p_t$ distributions
are expected for these background processes. 

To conclude, the exclusive cross sections of resonance production in 
ultraperipheral nuclear collisions are rather low but measurable at LHC 
for some particular channels and become larger at higher energies. 
The detailed analysis of other possibilities (especially, of collective effects 
in the magnetic field) should be done to get final conclusions.

\medskip

{\bf \large Acknowledgements}

\medskip

This work was supported by RFBR grant 18-02-40131 and
by the RAN-CERN program.\\


\begin{thebibliography}{99}
\bibitem{beg}
V.D. Burkert, L. Elouadrhiri, F.X. Girod, Nature {\bf 557} (2018) 396
\bibitem{ps}
M.K. Polyakov, P. Schweizer, arXiv:1812.06143
\bibitem{fss}
S. Ferreres-Sole, T. Sjostrand, Eur. Phys. J. C {\bf 78} (2018) 983
\bibitem{dr1}
I.M. Dremin, Physics {\bf 1} (2019) 33
\bibitem{hkr}
L.A. Harland-Lang, V.A. Khoze, M.G. Ryskin, Eur. Phys. J. C {\bf 79} (2019) 39
\bibitem{adl}
S.L. Adler, Ann. Phys. {\bf 67} (1971) 599
\bibitem{bms}
V.N. Baier, A.I. Milstein, R.Zh. Shaisultanov, Phys. Rev. Lett. {\bf 77} (1996)
1691
\bibitem{as}
S.L. Adler, C. Schubert, Phys. Rev. Lett. {\bf 77} (1996) 1695
\bibitem{bgms}
V.M. Budnev, I.F. Ginzburg, G.V. Meledin, V.G. Serbo, Phys. Rep. C {\bf 15} 
(1975) 181
\bibitem{kn}
S. Klein, J. Nystrand, Phys. Rev. Lett. {\bf 84} (2000) 2330
\bibitem{kmr}
V.A. Khoze, A.D. Martin, M.G. Ryskin, arXiv:1902.08136
\bibitem{deb}
R. Debbe, [STAR Collaboration] arXiv:1310.7044
\bibitem{star}
L. Adamczyk et al. [STAR Collaboration] Phys. Rev. C {\bf 96} (2017) 054904
\bibitem{kgs}
M. Klusek-Gawenda, A. Szczurek, arXiv:1609.04355
\bibitem{totem1}
G. Antchev et al. [TOTEM Collaboration] arXiv:1712.06153
\bibitem{dr2}
I.M. Dremin, Particles {\bf 2} (2019) 57
\bibitem{atl}
M. Aaboud et al. [ATLAS Collaboration] Nature Phys. {\bf 13} (2017) 852; 
Phys. Rev. D {\bf 99} (2019) 012008
\bibitem{cms}
D. d'Enterria [CMS Collaboration], arXiv:1808.03524 
\bibitem{ll}
L.D. Landau, E.M. Lifshitz, Phys. Zs. Sowjet. {\bf 6} (1934) 244
\bibitem{vyzh}
M.I. Vysotsky, E.V. Zhemchugov, arXiv:1806.07238
\bibitem{pdg}
PDG group. Review of particle physics. Phys. Rev. D {\bf 98} (2018) 42
\bibitem{khar}
D.E. Kharzeev, Prog. Part. Nucl. Phys. {\bf 75} (2014) 133

\end{thebibliography}
\end{document}